\documentclass[10pt,twocolumn,aps,pre,superscriptaddress]{revtex4-1}

\usepackage{amsmath}
\usepackage{graphicx}
\usepackage{color}
\usepackage{upgreek}
\usepackage[linkcolor = blue, citecolor = blue, urlcolor = blue, colorlinks = true]{hyperref}
\usepackage[version=3]{mhchem}
\usepackage{commath,amssymb}
\usepackage{ushort}
\usepackage{multirow}
\usepackage[usenames,dvipsnames]{xcolor}

\begin{document}

\author{Joost de Graaf}
\email{jgraaf@icp.uni-stuttgart.de}
\affiliation{Institute for Computational Physics (ICP), University of Stuttgart, Allmandring 3, 70569 Stuttgart, Germany}

\author{Arnold J.T.M. Mathijssen}
\affiliation{The Rudolf Peierls Centre for Theoretical Physics, 1 Keble Road, Oxford, OX1 3NP, United Kingdom}

\author{Marc Fabritius}
\affiliation{Institute for Computational Physics (ICP), University of Stuttgart, Allmandring 3, 70569 Stuttgart, Germany}

\author{Henri Menke}
\affiliation{Institute for Computational Physics (ICP), University of Stuttgart, Allmandring 3, 70569 Stuttgart, Germany}

\author{Christian Holm}
\affiliation{Institute for Computational Physics (ICP), University of Stuttgart, Allmandring 3, 70569 Stuttgart, Germany}

\author{Tyler N. Shendruk}
\affiliation{The Rudolf Peierls Centre for Theoretical Physics, 1 Keble Road, Oxford, OX1 3NP, United Kingdom}

\title{Understanding the Onset of Oscillatory Swimming in Microchannels}

\date{\today}

\begin{abstract}
Self-propelled colloids (swimmers) in confining geometries follow trajectories determined by hydrodynamic interactions with the bounding surfaces. However, typically these interactions are ignored or truncated to lowest order. We demonstrate that higher-order hydrodynamic moments cause rod-like swimmers to follow oscillatory trajectories in quiescent fluid between two parallel plates, using a combination of lattice-Boltzmann simulations and far-field calculations. This behavior occurs even far from the confining walls and does not require lubrication results. We show that a swimmer's hydrodynamic quadrupole moment is crucial to the onset of the oscillatory trajectories. This insight allows us to develop a simple model for the dynamics near the channel center based on these higher hydrodynamic moments, and suggests opportunities for trajectory-based experimental characterization of swimmers' hydrodynamic properties.
\end{abstract}

\maketitle

The locomotion of self-propelled particles (swimmers) typically occurs at boundaries or under confinement. Accurately describing the effect of confinement on swimmers is therefore of significant interest to understanding the behavior of microorganisms and artificial swimmers. In modelling these systems, hydrodynamic interactions (HIs) are often ignored, which is a valid approximation in some cases, such as when microbial swimmers' run-and-tumble dynamics dominate~\cite{mathijssen2015hotspots}. However, HIs can play an important role, e.g., see Refs.~\cite{Lauga06,Drescher11,Leonardo11,Jana12,Lushi014}, and therefore cannot be \textit{a priori} ignored in modelling. Recent experiments on self-phoretic colloidal swimmers have shown that their orientation is strongly influenced by HIs due to the presence of a wall~\cite{das15}. However, there is ongoing debate on the importance of near-wall effects and the level at which to truncate the hydrodynamic moment expansion~\cite{spagnolie12}.
 
A specific example of this are the helical and oscillatory trajectories of single swimmers in confining geometries as observed experimentally by Jana~\textit{et al.}~\cite{Jana12} and in simulations~\cite{zhu13,wu15,shum15}. Such oscillatory trajectories appear to be common place, having been reproduced by many models, and independent of specific swimmer type. However, a physical understanding of these oscillations remains wanting. It is indisputable that the oscillations do not arise simply from the lowest order hydrodynamics, which result in direct attraction to surfaces~\cite{berke08}, while the inclusion of higher-order modes can lead to more complex behavior~\cite{spagnolie12,Ishimoto13}. Although observed in confined quiescent fluids, these oscillatory trajectories are reminiscent of those observed in the rheotaxis of swimmers subjected to external flows~\cite{zoettl12}, which result primarily from the interplay between the flow and persistent particle motion due to self-propulsion. Z{\"o}ttl and Stark indicate that near-field lubrication theory can be used when there is no externally applied flow to describe such trajectories~\cite{zoettl14,rusconi14}. Yet, the observations of Zhu~\textit{et al.} demonstrate that oscillatory trajectories arise in a channel that is three times as wide as the self-propelled particle. Additionally, the trajectories of squirmers close to single walls in quiescent fluid show oscillations~\cite{Li14}, which have been explained by the competition between far-field HIs and short-ranged wall-swimmer potentials~\cite{Lintuvuori15}. Thus, there is a clear need to establish to what extent the observation of oscillatory trajectories in systems with confinement originate from a near- or far-field effect and, in conjunction, to assess the importance of higher-order hydrodynamic modes. 

\begin{figure}[ht] 
\centering 
  \includegraphics[width=0.85\linewidth]{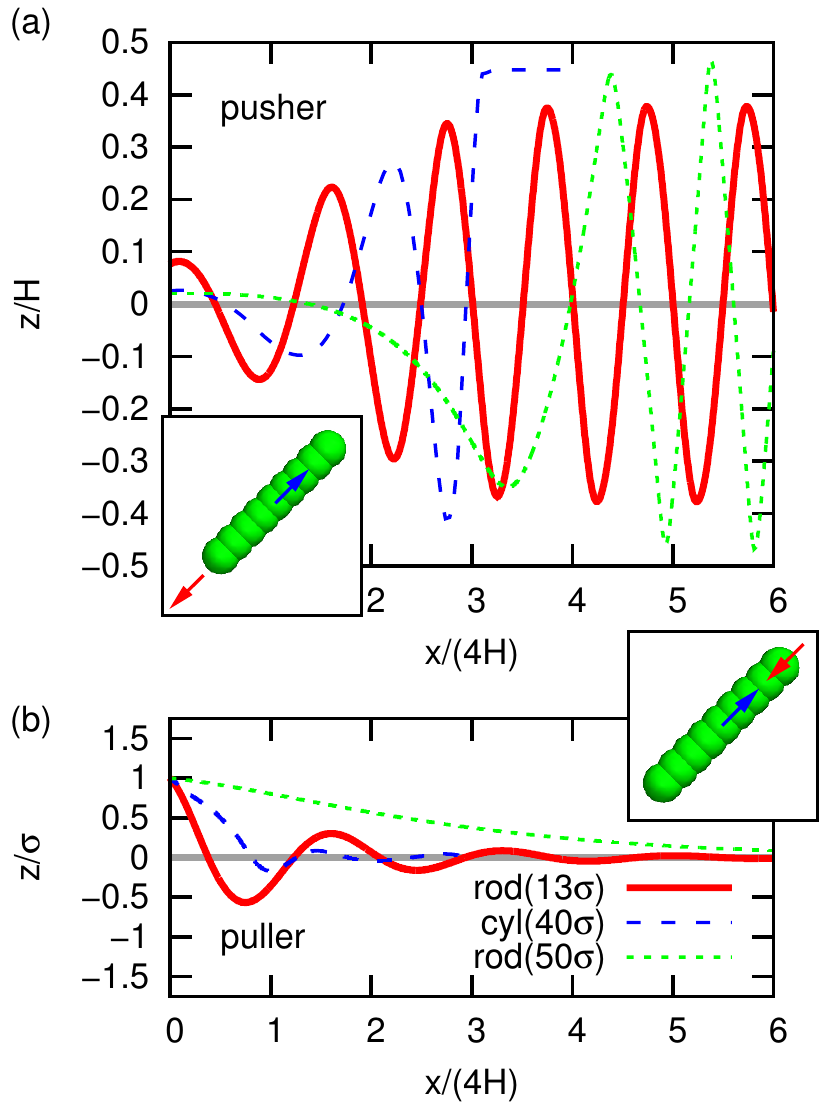} 
  \caption{The trajectory of swimmers between two parallel plates with separation $H$. The horizontal displacement $x$ and vertical position $z$ are given for a swimmer that is initially oriented parallel to the walls and at $z=1\sigma$, with $\sigma$ the MD unit of length (LB lattice size) and $z=0$ the center of the channel. (a) The results for pushers: rod for $H=13\sigma$ (red, solid), cylinder for $H=40\sigma$ (blue, dashed), and rod for $H=50\sigma$ (green, dots). The inset shows a schematic representation of the raspberry rod-swimmer (scaled for $H=13\sigma$), the force is indicated in red and counter force in blue. (b) The results for pullers, otherwise the systems are the same. Appendix~\ref{sub:fig1} contains a companion figure showing the evolution of the angle $\phi$ for these swimmers.} 
  \label{fgr:traj} 
\end{figure} 
 
In this manuscript, we demonstrate that the onset of time-varying oscillatory trajectories in systems confined within a channel and without external flow can be well-understood using far-field theory~\cite{mathijssen16}. We investigate the specific case of two parallel infinite plates that enclose the fluid and a single rod-shaped swimmer, using our lattice-Boltzmann (LB) `raspberry' force/counter-force formalism~\cite{degraaf16a} (Fig.~\ref{fgr:traj}; insets). We have previously shown that the rod-shaped LB swimmers have well-defined higher-order hydrodynamic moments~\cite{degraaf16a}; see Table~\ref{table:multipoleMoments} for representations of the first five moments. These simulations conclusively show that a puller-type rod that starts far from the wall but off-center follows a damped oscillatory trajectory towards the middle of the channel, whereas a pusher-type rod moves between the walls along a sinusoidal path with increasing amplitude. Surprisingly, the oscillations are observable even for plate separations as great as ten times the length of the rod. We explain these observations within the framework of our far-field hydrodynamic theory: the dipole and octupole moments induce hydrodynamic forces towards the center (puller) or towards the walls (pusher), while the quadrupole moment causes pure oscillatory motion. The oscillatory trajectories within plate confinement thus provides an indirect means to characterize the hydrodynamic properties of swimmers, which would grant access to moments beyond those that can be obtained from lattice swimming~\cite{brown16} or tracer paths~\cite{degraaf16a}.
 
We consider two raspberry swimmers (rod and cylinder) in the main text to study the movement of shape-anisotropic swimmers under confinement using our \textsf{ESPResSo} LB implementation~\cite{degraaf16a,arnold13a}. Their construction and characterization in terms of hydrodynamic moments, as well as the fluid and coupling parameters, are introduced in Refs.~\cite{fischer15,degraaf16a} and detailed in Appendix~\ref{sub:model}. Our swimming model's essential aspect is that a force is applied to the body, consisting of many fluid-particle coupling points, and the system is made force free by applying an equal and opposite force to the fluid, see the insets in Fig.~\ref{fgr:traj}. This coupling gives rise to a series of hydrodynamic modes for anisotropic particles~\cite{degraaf16a}. 

These raspberry particles are placed in an LB fluid between two parallel (no-slip) bounce-back plates, with normals in the $\hat{z}$ direction, separated by a distance $H$. The fluid domain is periodic in the other two ($xy$) directions. The vertical position of the swimmer's center of mass (CM) is indicated using $z \in [-H/2,H/2]$, with $z=0$ the middle of the channel. Lateral displacement is given by $x$ and measured from the swimmer's initial position ($x=0$) --- our trajectories are straight in the $xy$-plane. Finally, the angle of the swimmer's director $\hat{p}$ (which points along the main axis) with the plate normal $\hat{z}$ is given by $\phi \in [-\pi/2,\pi/2]$, with $\phi = 0$ swimming parallel to the plates. To prevent the swimmers from penetrating the wall, we imposed a short-ranged (almost hard) Weeks-Chandler-Anderson (WCA) interaction between the raspberry swimmers and the walls (Appendix~\ref{sub:setup}). This wall-swimmer interaction is necessary as our LB algorithm does not explicitly account for near-wall lubrication corrections~\cite{fischer15}. All of the results shown in the main text employ a WCA diameter $d = \sigma$, with $\sigma$ the LB lattice size. We limit the swimming speed to ensure the low Reynolds number regime, $Re < 0.01$.

Figure~\ref{fgr:traj} and the supplemental movies (not included in arxiv version) demonstrate the onset of oscillatory trajectories. These are representative sample swimmer trajectories, where the swimmers start off-center and oriented parallel to the plates. Both the rod and cylinder models of pushers and pullers display time-varying oscillatory behavior. In the specific case of our rods, the wavelength of the oscillations is $\lambda \approx 4H$. All pushers move towards the wall and the pullers move towards the center of the plates. After only a few periods, these pullers move along the centreline of the channel and these pushers have arrived in the near-wall region, where swimmer specific details and lubrication corrections would be required to accurately predict dynamics. Oscillations are observed for all cylinder and rod swimmers in plate separations that we could simulate ($5\sigma \le H \le 50\sigma$). The rod is $\sim5\sigma$ in length, thus the oscillatory trajectories arise in systems with a channel height to particle size ratio up to at least 10.

To verify the generality of the initial oscillations, we considered several initial positions $z$ and orientations $\phi$ for rod pusher and puller swimmers. We found that depending on the type of swimmer and its initial position/orientation, several oscillations in the physical regime can be observed, before near-wall effects cannot be ignored. We further showed that oscillations for rod-like swimmers appear for a large range in rod aspect ratios (Appendix~\ref{sub:length}). At long times the LB pusher rods display a limit cycle, whereas the pusher cylinder does not. To what extent such a limit cycle (Fig.~\ref{fgr:traj}a; solid red curve) or sliding dynamics (Fig.~\ref{fgr:traj}a; dashed blue curve) might be physical is not considered here, as algorithmic artifacts close to the walls impact the near-wall dynamics. Appendix~\ref{sub:limit} provides a discussion of these limitations and this work does not confirm their physical nature~\cite{shum15}. However, our results establish that the initial oscillations before the rod comes close to the wall (a proximity of $\sim 2 \sigma$) are physical. It is this onset of oscillatory trajectories that we concern ourselves with here and subject to theoretical analysis in the following.
 
We model the raspberry swimmers theoretically as ellipsoids with aspect ratio $\gamma$, position $\vec{r}$ and orientation $\hat{p} = (\cos \phi, 0, \sin \phi)$. Due to its motion, the swimmer generates a flow field $\vec{u}$, which we define in terms of a multipole expansion of the Stokeslet flow solution. Spagnolie~\textit{et al.} argue that far-field HIs give surprisingly accurate results, when compared to theory that includes a finite-size correction to more accurately account for near-field effects, even for small swimmer-wall separations~\cite{spagnolie12}. Hence, the flow at position $\vec{x}$ generated by the force-free and torque-free swimmer is 
\begin{align} 
  \label{eq:swimmerVelocityDefm} 
  \vec{u}\left(\vec{x}, \vec{r}, \hat{p} \right) &=  
                        \kappa \vec{u}_\textmd{D} +  
                           \nu \vec{u}_\textmd{Q} +  
                          \mu \vec{u}_\textmd{SD} +  
                       o_1 \vec{u}_{\textmd{O}_1} +  
                       o_2 \vec{u}_{\textmd{O}_2} +  
                                             \ldots 
\end{align} 
Here, $\vec{u}_\textmd{D}$ is the Stokes dipole that models the force balance between propulsion and drag, $\vec{u}_\textmd{Q}$ is the quadrupole that represents the fore-aft asymmetry of the propulsion mechanism, $\vec{u}_\textmd{SD}$ is the quadrupolar source doublet that is associated with the finite size of the swimmer, and $\vec{u}_{\textmd{O}_1}$ and $\vec{u}_{\textmd{O}_2}$ are the two octupolar terms (Appendix~\ref{sub:model}). The shape of these moments in bulk is shown in Table~\ref{table:multipoleMoments}. Note that this is a point-based expansion, which should not be confused with the squirmer expansion for finite-sized spheres; in the far-field these expansions can be mapped onto each other.
 
The effect of the confining walls (two parallel no-slip plates) is now accounted for by the method of images, where we truncate the approximation after four image systems on each side of the microchannel. Subsequently, the flow wall-induced flow advects and reorients the swimmer according to the Fax{\'e}n relations, resulting in the translational and angular velocities $\vec{v}_\textmd{HI}$ and $\vec{\Omega}_\textmd{HI}$ (Appendix~\ref{sub:images}). In Ref.~\cite{mathijssen16} details are given of the procedure by which to obtain these velocities in terms of the multipole coefficients. The swimmer's equations of motion are given by 
\begin{eqnarray} 
  \label{eq:EquationsOfMotion}
  \dot{\vec{r}} = v_s \vec{p}  + \vec{v}_\textmd{HI}, \quad  
  \dot{\hat{p}} = \vec{\Omega}_\textmd{HI} \times \hat{p}, 
\end{eqnarray} 
where $v_s$ is the autonomous swimming speed, and the velocities $\vec{v}_\textmd{HI}$ and $\vec{\Omega}_\textmd{HI}$ are functions of the multipole coefficients. 
 
\begin{table}[t] 
  \begin{tabular}{| l || r | r || r | r |} 
  \hline 
                                   &   \multicolumn{2}{c||}{Rod}    &   \multicolumn{2}{c|}{Cylinder}   \\   
  \hline 
  $\Downarrow$ Coefficient         &              LB &       theory &                LB &        Theory \\  
  \hline  
  \hline 
                          $\kappa$ &     $\pm 0.013$ &  $\pm 0.0153$ &       $\pm 0.027$ & $\pm 0.0312$ \\ 
                             $\nu$ &         $0.038$ &      $0.0294$ &            $0.21$ &      $0.194$ \\ 
                             $\mu$ &           $0.0$ &         $0.0$ &             $0.0$ &        $0.0$ \\ 
                             $o_1$ &           $0.0$ &         $0.0$ &             $0.0$ &        $0.0$ \\ 
                             $o_2$ &     $\pm 0.113$ &  $\pm 0.1256$ &        $\pm 2.11$ &  $\pm 2.176$ \\ 
                             $v_s$ &       $ 0.0025$ &      $0.0025$ &         $ 0.0010$ &    $ 0.0010$ \\ 
  \hline 
  \multicolumn{5}{c}{\includegraphics[width=0.95\linewidth]{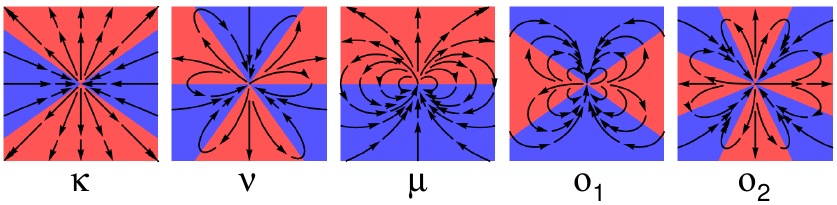}} \\
  \end{tabular} 
  \caption{Multipole moments of the swimmer-generated flow field for the two swimmer types: the rod and the cylinder. The columns labelled `LB' provide the values measured in our previous study by means of Legendre-Fourier decomposition in a close-to-bulk system with periodic boundary conditions~\cite{degraaf16a}. The columns labelled `theory' provide the moments fitted from the trajectory in our confining geometry by using the theoretical model~\eqref{eq:EquationsOfMotion}. Values are given in LB simulation units, and the positive/negative signs correspond to pusher and puller swimmers, respectively. The bottom row shows representations of the flow field of the first five hydrodynamic moments in bulk: dipole $\kappa$, quadrupole $\nu$, source dipole $\mu$, source octupole $\textmd{o}_1$, and octupole $\textmd{o}_2$. The arrows are stream lines and the colors indicate flow away from (red) or towards (blue) the swimmer.}
  \label{table:multipoleMoments} 
\end{table} 

To predict the swimmer dynamics of Fig.~\ref{fgr:traj} theoretically, we integrate the equations of motion~\eqref{eq:EquationsOfMotion}. Here, we use the same swimming speed and initial conditions as in the LB simulations, but we allow the multipole coefficients to vary about their measured values. We can thus fit the multipole coefficients via a least-squares method. To obtain the best agreement with the measured trajectory, we used the four initial oscillations (Appendix~\ref{sub:fitting}). The 3\textsuperscript{rd} and 5\textsuperscript{th} columns of Table~\ref{table:multipoleMoments} show the multipole coefficients found in this manner for swimmers of the rod and cylinder type, respectively. Using only a single oscillation leads to a change in the fitted values of $\sim 20\%$, showing our method to be robust and requiring only fragments of a trajectory to be effective. In addition, we verified that the result of the fitting was independent of the height $H$ of the channel, eliminating the possibility of boundary artifacts. In our previous work~\cite{degraaf16a}, we obtained the multipole coefficients directly from the flow field of the swimmers in our LB simulations by means of projection via a Legendre-Fourier decomposition. These values are listed in the 2\textsuperscript{nd} and 4\textsuperscript{th} columns of Table~\ref{table:multipoleMoments}, respectively. The projection was carried out in the absence of confinement, using a large simulation box with periodic boundary conditions, for which the finite-size effects differ strongly from those of the confining geometry. There is excellent correspondence between the two measurements of the hydrodynamic moments for both swimmer shapes. This demonstrates the applicability of far-field theory to describe the onset of the observed oscillatory trajectories. The far-field result is accurate until the swimmer-wall distance becomes too small.

Let us now focus on the general features of the theoretical model and analyze the impact of the various hydrodynamic moments on the motion of the swimmer. Firstly, our calculations confirm that the pusher swimmer ($\kappa > 0$) undergoes oscillatory trajectories that move away from the center of the channel, and pullers ($\kappa < 0$) converge towards the centerline. However, oscillations about the center only occur if the quadrupolar terms are included, and the oscillation wavelength decreases with the associated quadrupolar coefficients $\nu$ and $\mu$. A spherical swimmer with $\nu = \mu = 0$~\cite{degraaf16a} does not display such oscillations. The octupolar contributions further control the damping and growth of the trajectories, where the positive signs of $o_1$ and $o_2$ correspond to motion towards the boundaries. The aspect ratio $\gamma$ leads only to a second-order correction in the theory. That is, the hydrodynamic moments dominate the dynamics of the swimmer, therefore rods with different aspect ratios still show similar oscillations (Appendix~\ref{sub:length}).
 
\begin{figure}[t] 
  \includegraphics[width=\linewidth]{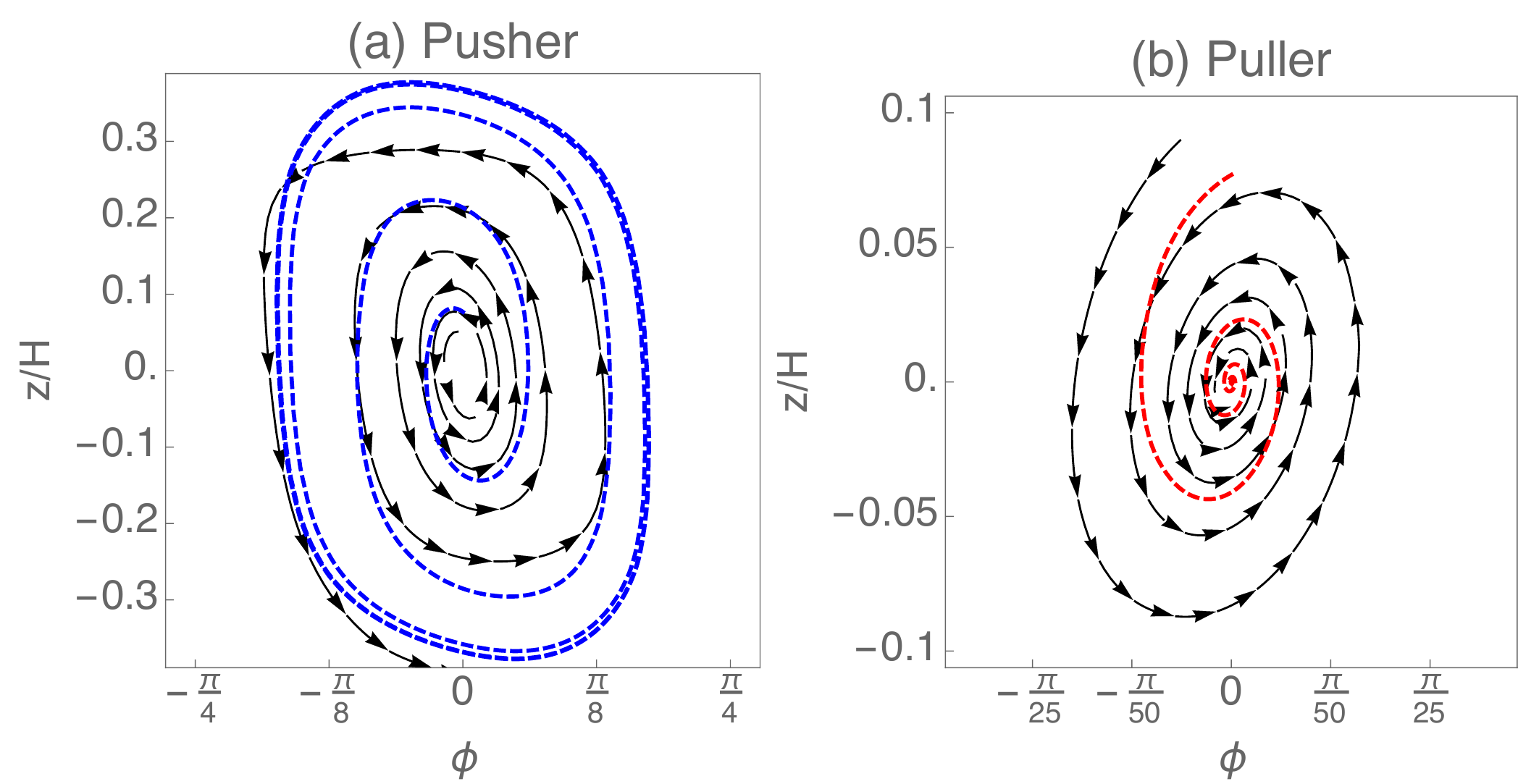} 
  \caption{Trajectories of swimmers in ($\phi$,$z$) phase space, for a rod-type pusher (a) and puller (b) with a wall separation of $H = 13 \sigma$. The LB simulation data are shown as thick, dashed, blue and red lines. The theoretical predictions are superimposed as black arrows.} 
  \label{fgr:phaseSpace} 
\end{figure} 
 
The dynamical system can be understood further by considering the motion of the swimmer in phase space. Due to the translational invariance in the $x$ and $y$ coordinates, the equations of motion can be reduced to two coupled first-order PDEs in ($\phi$,$z$) space, next to the uncoupled equation for the $x$ coordinate. Figure~\ref{fgr:phaseSpace} shows the LB swimmer trajectories in phase space, superimposed with the theoretical model, where the fitted multipole moments in Table~\ref{table:multipoleMoments} have been used. The dipolar term leads to a star-type fixed point (curves radiating from a point) at the origin, that is stable for pullers and unstable for pushers. The oscillatory motion due to the quadrupolar contributions corresponds to a circle-type phase-space trajectory (closed loops around a point) centered on the origin. Together the dipole and quadrupole produce a spiral. For pushers, the trajectories spiral outwards (Fig.~\ref{fgr:phaseSpace}a), and inwards for pullers (Fig.~\ref{fgr:phaseSpace}b). The theoretical predictions do not show a limit cycle in Fig.~\ref{fgr:phaseSpace}. Both the far-field framework and the LB method are unable to adequately capture hydrodynamic interactions in the near-wall region and further study of this region, where both lubrication corrections~\cite{zoettl14} and short-ranged potentials~\cite{Lintuvuori15} can play a role, is required. 

Our result shows that movement of a swimmer under confinement can in principle be used to quantitatively determine the hydrodynamic moments, even up to the octupolar moment as shown here. Specifically, about one period is the minimum path length required to fit these modes to within 20\%. This suggests that our method has applicability to experimental systems where thermal noise and tumbling can effect the trajectory. The presence of these sources of noise would require ensemble averaging trajectories in $(\phi,z)$ space, which can then be fitted using our procedure. Noise also implies that parts of the trajectory will occur many times during measurements, meaning the near-wall dynamics in which we observed a limit cycle, does not play an important role. One simply averages many different trajectories away from the wall to improve the fitting statistics.

\begin{figure}[t] 
\centering 
 \includegraphics[width=\linewidth]{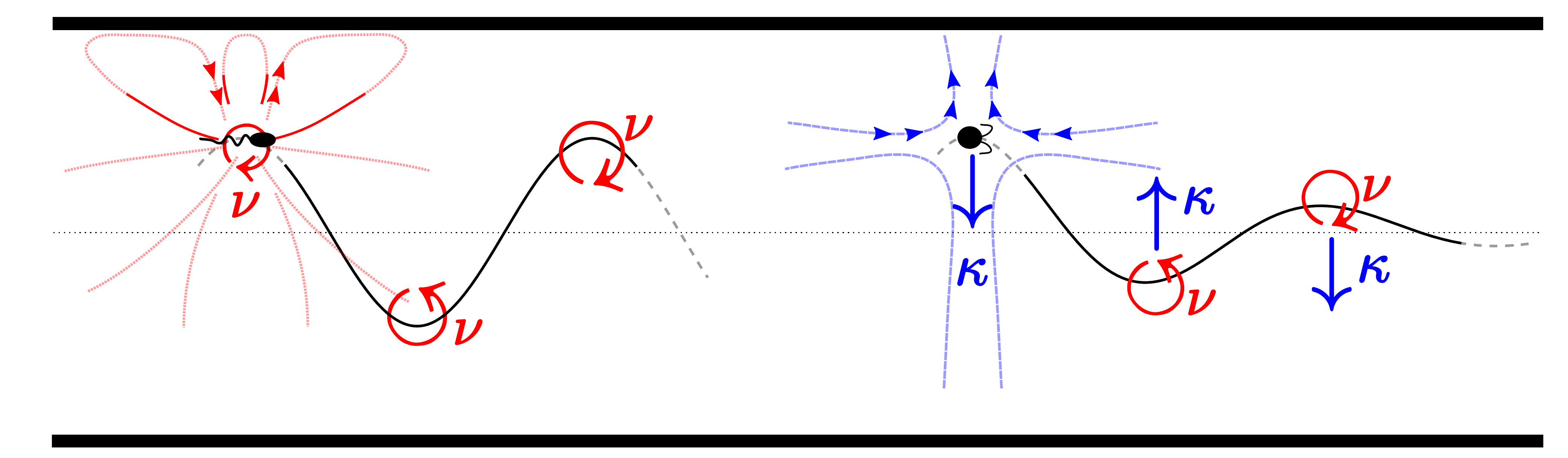} 
 \caption{Illustration of the mechanisms of oscillatory swimmer motion in microchannels. (a) Quadrupolar moment only. The HI rotates the swimmer away from the nearest wall resulting in an oscillatory trajectory. (b) Quadrupole and puller-type dipole. The dipole pushes the swimmer away from the nearest wall, decreasing the oscillation amplitude.} 
 \label{fgr:cartoon} 
\end{figure} 

A physical intuition for the onset of oscillatory swimming can be distilled from the LB simulations and far-field hydrodynamic description by considering the trajectory of a swimmer initially set at $z_0$ near the centerline and oriented parallel to the walls (Fig.~\ref{fgr:cartoon}). Since our raspberry swimmers have large quadrupolar moments ($\nu\sim10^{-1}$), we first consider only the flow fields generated by the positive quadrupole. This flow serves to rotate the swimmer away from the nearest wall, as shown schematically in Fig.~\ref{fgr:cartoon}a. The continual rotation away from the nearest wall establishes the oscillations. By linearly expanding the equations of motion \eqref{eq:EquationsOfMotion} about the centerline, the micro-swimmer dynamics can be captured by a linear system of coupled differential equations (Appendix~\ref{sub:oscillations}). Whenever there is only a quadrupolar flow field, the trajectory is approximated to be simple oscillatory motion $z\left(t\right) \approx z_0 \cos\left( \omega t \right)$ with angular frequency $\omega = 4\left( 3 \nu v_s / H^{5}\right)^{1/2} $ and wavelength $\lambda \approx 2\pi v_s/\omega$. Although $\mu=0$ in this study, a source dipole moment also leads to simple oscillations (Appendix~\ref{sub:oscillations}). This also theoretically explains the observations of persistent oscillations for neutral squirmers made by Zhu~\textit{et al.}~\cite{zhu13}, even though there are differences in the confining geometry. Next we add the dipolar term to the expansion
\begin{align} 
  z\left(t\right) &\approx z_0 e^{ \alpha t } \cos\left( \omega t \right) 
  \label{eq:simple} 
\end{align} 
where $\alpha=3\kappa/H^3$, which is negative for pullers. The dipolar term also modifies the frequency $\omega$ due the wall-induced rotation $\vec{\Omega}_\textmd{HI}$, but this effect is negligible if $\nu \gg 81 \kappa^2 / \left( 48 H v_s \right)$, which is the case here. A pusher also obeys equation \eqref{eq:simple} but with $\alpha>0$ and exponentially growing amplitudes, which leads to a rapid breakdown of the near-centerline assumption. The sensitivity of oscillations to channel height is unmistakable in the $\exp\left(H^{-3}\right)$-dependence of \eqref{eq:simple} reflecting the fact that the essential hydrodynamic moments are high order. Whereas higher order moments are required to predict the oscillation wavelength and damping factor quantitatively, the dipolar and quadrupolar moments can be fit from the dynamics using \eqref{eq:simple} with a error margin of $\sim40\%$ compared to the LB-measured values. Hence, \eqref{eq:swimmerVelocityDefm}-\eqref{eq:simple} allow for characterization of the swimmer's hydrodynamic properties based on experimental trajectories and can be readily transferred to the observations made by Zhu~\textit{et al.}~\cite{zhu13}. Likewise, LB raspberry simulations can be extended to more complex 3D geometries such as square channels and round tubes, in which we observed helical motion (Appendix~\ref{sub:helical}).
 
In conclusion, we have studied the onset of oscillatory motion of swimmers in microchannels without externally applied flow and in an otherwise quiescent medium using both LB simulations and hydrodynamic theory. The pusher-type swimmers follow a sinusoidal trajectory with increasing amplitude, whereas pullers perform a damped oscillation towards the center of the channel. Our results and previous observations of such phenomena~\cite{zhu13} can be explained by our theoretical model, which uses far-field hydrodynamics only. We conclude that the onset of oscillations can be described without taking into account near-wall lubrication effects as has been previously presumed~\cite{zoettl14} provided that a quadrupole moment (or source-dipole) is accounted for in addition to the primary dipole moment. To fully characterize particle trajectories in relatively wide channels, many hydrodynamic moments are required, as high as the octupole in our case. However, the excellent match of our trajectory-fitted moments to those measured in bulk suggests that similar experimental measurements can be used to determine the hydrodynamic moment decomposition of microorganisms or artificial swimmers. Our work stresses the relevance of far-field hydrodynamics in confining geometries and thus opens the way for new studies that aim to exploit these insights in microfluidic environments. Future work will focus on the analysis of more complex force/counter-force swimmers to model the richness in shape and hydrodynamic moments of experimentally available swimmers.
 
\textit{Acknowledgements} --- AJTMM and TNS gratefully acknowledge funding from the ERC Advanced Grant (291234 MiCE) and EMBO (ALTF181-2013); JdG from an NWO Rubicon Grant (\#680501210) and a Marie Sk{\l}odowska-Curie Intra European Fellowship (G.A. No. 654916) within Horizon 2020. JdG and CH further thank the DFG for funding through the SPP 1726 ``Microswimmers --- From Single Particle Motion to Collective Behaviour''. We would also like to thank A. Z{\"o}ttl, A. Doostmohammadi, and A.T. Brown for useful discussions.

\appendix

\section{\label{sec:method}Simulation Details}

In this section we present details of the simulation model that complement the description given in the main text. In addition, we provide further simulation results, which serve to underpin the generality of our findings.

\subsection{\label{sub:model}Fluid Parameters and Swimmer Models}

The `raspberry swimmers' are based on the lattice-Boltzmann method implementation~\cite{degraaf16a} and simulated using a graphics processing unit (GPU) based LB solver~\cite{roehm12} that is attached to the MD software \textsf{ESPResSo}~\cite{limbach06a,arnold13a}. This GPU LB employs a D3Q19 lattice and a fluctuating multi-relaxation time (MRT) collision operator~\cite{dhumieres02}. All of our simulations are performed in a quiescent (unthermalized) LB fluid. A three-point interpolation stencil~\cite{ladd94} is employed together with the LB viscous coupling of Ref.~\cite{ahlrichs99} to couple the raspberry particles to the fluid. We set the fluid density to $\rho = 1.0 m_{0}\sigma^{-3}$, the lattice spacing to $1.0\sigma$, the time step to $\Delta t = 0.005 \tau$ ($\tau$ is the time and $m_{0}$ the mass unit), the (kinematic) viscosity to $\nu = 1.0 \sigma^{2}\tau^{-1}$, and the bare particle-fluid friction to $\zeta_{0} = 25 m_{0}\tau^{-1}$. Fischer~\textit{et al.}~\cite{fischer15} provide a detailed description of the dimensionless numbers that specify the fluid properties to which these choices correspond.

\begin{figure}[!htb]
  \centering
  \includegraphics[scale=1.0]{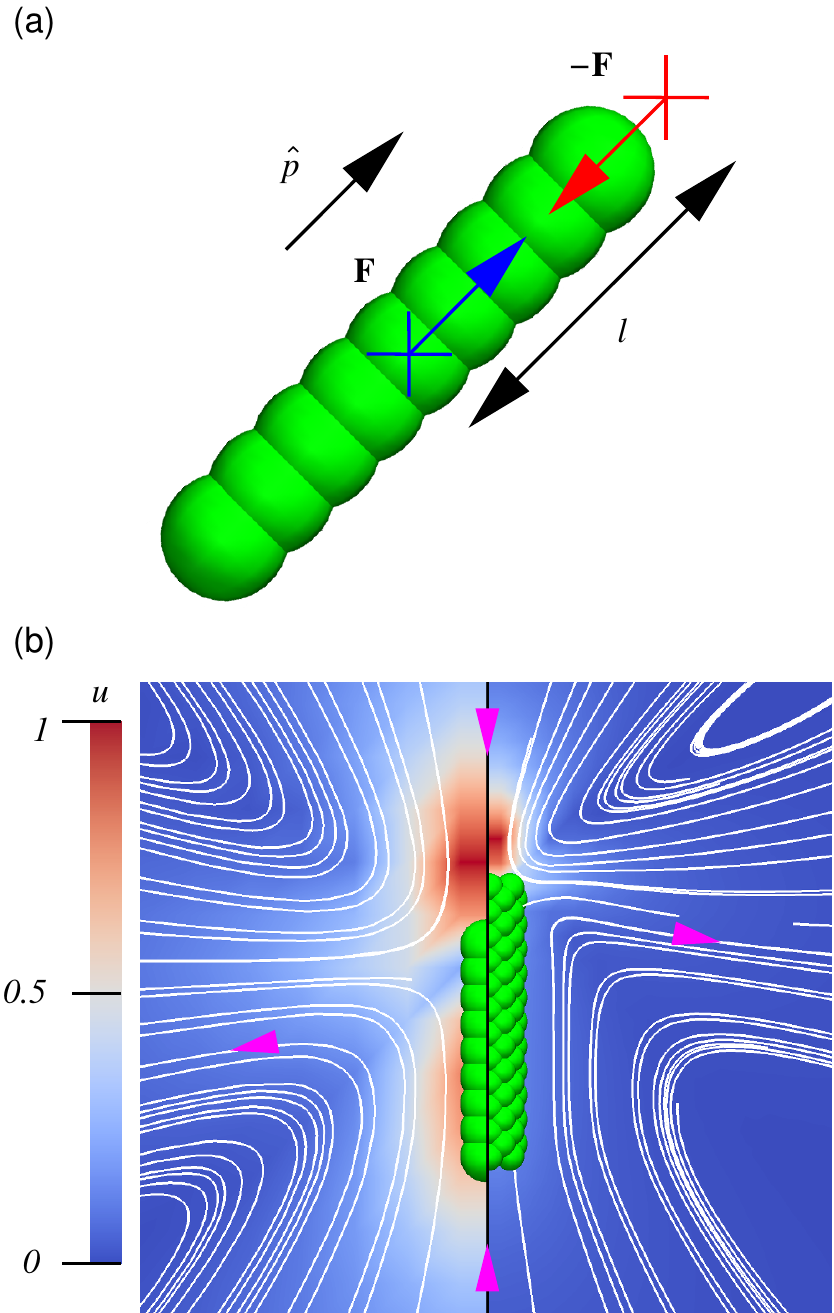}
  \caption{\label{fig:rasp}The flow field around our raspberry-swimmer models. (a) Sketch of a puller-type rod. The size of the green spheres roughly corresponds to the effective hydrodynamic radius of our coupling points ($\sim0.5\sigma$). A force $\vec{F}$ (blue arrow) is applied to the central bead (blue cross) in the direction of the symmetry axis $\hat{p}$ (black arrow). A counter force $-\vec{F}$ (red arrow) is applied to the fluid at a point $l \hat{p}$ (red cross), with $l$ the separation length. (b) The flow field around a puller-type raspberry rod (left) and cylinder (right). The normalized magnitude of the flow velocity in the lab frame given by the legend (red $\max \vert \vec{u}(\vec{r}) \vert = 1$, dark blue $\vert \vec{u}(\vec{r}) \vert = 0$); only a part of the simulation box is shown, again the diameter of the green spheres is $1\sigma$. White curves are stream lines and the magenta arrow heads indicate the direction of flow.}
\end{figure}

We consider two types of self-propelled particles, a rod and cylinder as shown in Fig.~\ref{fig:rasp}. The rod consists of nine coupling points spaced $0.5\sigma$ apart over a line, with $\sigma$ the LB grid spacing. The cylinder consists of $161$ coupling points spread over $23$ groups of hexagonal disks (seven particles with distance $\sigma$), stacked alternatingly with a separation of $0.5\sigma$ along the axis. The rod has an effective hydrodynamic length of 5.5 and a diameter of 1.7; the cylinder has an effective length of 12 and diameter of 3.2; and the sphere an effective radius of 3.1~\cite{degraaf16a}. Full details of these swimmers construction (mass, rotational inertia, etc.) are given in Ref.~\cite{degraaf16a}. 

The raspberry bodies are made into swimmers by assigning a unit (direction) vector $\hat{p}$ to their center that points along the symmetry axis, see Fig.~\ref{fig:rasp}a. This $\hat{p}$ co-moves with the particle. We apply a force $\vec{F}$ in the direction of $\hat{p}$ ($\vec{F} = F \hat{p}$) to the central molecular-dynamics bead, to which the rest of the coupling points are rigidly attached. This force causes the raspberry particle to move. We further apply a counter force $-\vec{F}$ to the fluid at a position $l\hat{p}$, with $l$ the separation length, to make the system force free. For positive values of $l$ the swimmer is a puller and for negative values it is a pusher. We refer to Ref.~\cite{degraaf16a} for the specific parameter choices. For convenience, we summarize the relevant quantities that these choices lead to in Table~\ref{tab:active}, namely: the speed and hydrodynamic moments.

\begin{table}
  \begin{ruledtabular}
  \begin{tabular}{c|c|c|c|c|c|c}
  shape                      &      $v_{s}$ &        $\kappa$ &            $\nu$ & $\mu$ &  $o_{1}$ & $o_{2}$ \\
  \hline

  \multirow{2}{*}{rod}       & $2.5\;10^{-3}$ & $-1.3\;10^{-2}$ & $3.7\;10^{-2}$ &   0.0 &      0.0 & $-0.11$ \\
                             & $2.5\;10^{-3}$ &  $1.3\;10^{-2}$ & $3.7\;10^{-2}$ &   0.0 &      0.0 &  $0.11$ \\
  \hline

  \multirow{2}{*}{cylinder}  & $9.9\;10^{-4}$ & $-2.7\;10^{-2}$ &         $0.23$ &   0.0 &      0.0 &  $-2.1$ \\
                             & $1.0\;10^{-3}$ &  $2.7\;10^{-2}$ &         $0.23$ &   0.0 &      0.0 &   $2.1$ \\
  \end{tabular}
  \end{ruledtabular}
  \caption{\label{tab:active}The properties of our LB raspberry swimmers from Legendre-Fourier decomposition~\cite{degraaf16a}. The table provides the shape, the velocity $\nu_{s}$ of the swimmer in units of ($\sigma/\tau$), the dipole strength $\kappa$ ($\sigma^{3}/ \tau$), the quadrupole strength $\nu$ ($\sigma^{4}/ \tau$), the source-dipole strength $\mu$ ($\sigma^{4}/ \tau$), the source octupole $o_{1}$ ($\sigma^{5}/\tau$), and the force octupole $o_{2}$ ($\sigma^{5}/\tau$), respectively. The positive signs of $\kappa$ correspond to pusher swimmers and the negative ones to pullers.}
\end{table}

Using the size and speed of the swimmers, and kinematic viscosity of fluid, it is clear that Reynolds number of all our swimmers is less than 0.01. We use a quiescent fluid, therefore the P{\'e}clet number is ill-defined, as there is no translational (or rotational) diffusion. Both rod- and cylinder-type swimmers model `cylindrical' self-propelled particles, but the level of description varies as well as the speed of the simulation. The rod-like model captures some of the hydrodynamic aspects of an extended object, namely the existence of a hydrodynamic quadrupole. The use of the low number of coupling points makes the simulations fast compared to those for the cylinder swimmer. However, extended objects with a higher coupling-point density, like our cylindrical swimmer, more accurately model a rod-like shape that is impenetrable to the fluid~\cite{degraaf15a}. The use of a cylindrical swimmer thus serves to verify that the results obtained for the rod swimmer are not induced by low coupling-point density.

\subsection{\label{sub:setup}Simulation Setup}

The above LB and raspberry coupling parameters result in faithful reproduction of theoretical results for passive particles in confining geometries, as was shown by De Graaf~\textit{et al.}~\cite{degraaf15b}. Since we use a three-point coupling stencil deviations from the expected behavior of passive particles (solutions to the Stokes' equations) will occur within $2\sigma$ of the wall, rather than the $1\sigma$ found in Ref.~\cite{degraaf15b}. Here, confinement is achieved by placing two no-slip (bounce-back) walls on either side of the simulation box in the $z$-direction. We pad the box using two lattice nodes of wall (zero velocity) on either side rather than one, because of the 3-point coupling of our swimmers to the fluid. The simulation domain is kept periodic in the other ($xy$) directions. This leads to a so-called `slit-pore' geometry. We performed Poiseuille flow experiments to verify the height of the channel, the results of which we fit to the Hagen-Poiseuille expression. In all cases, the deviation between the imposed and fitted channel height is minimal ($\sim 0.1 \sigma$).

In each of our simulations, the swimmer is initialized in the center of the box in the $xy$-direction, at a height $z$ with respect to the center of the channel ($z=0)$. We ensure that the swimmer's director $\hat{p}$ is in the $xz$-plane and impose its initial angle $\phi$ with the plate normal $\hat{z}$, where $\phi \in [-\pi/2,\pi/2]$. Typically, we use $\phi = 0$ as the initial angle, which means that the swimmer is oriented parallel to the plane. The fluid in the channel is fully quiescent at time $t = 0$ and the particle starts with zero velocity. The LB parameters are chosen such that after the particle has moved only a fraction of $\sigma$ the fluid flow field and terminal velocity of the swimmer is established, thereby minimizing the effect of inertia and momentum-transport retardation that physically do not play a role on the colloidal length scale at low Reynolds number.

To prevent the swimmers from penetrating the wall, we include a Weeks-Chandler-Anderson (WCA) interaction between the raspberry coupling points and the bounce-back boundaries. The expression for the interaction is given by
\begin{align}
  \label{eq:WCA} U_{\mathrm{WCA}}(r) &= \left\{ \begin{array}{rc} 4 \epsilon \left[ \left( \frac{d}{r} \right)^{12} - \left( \frac{d}{r} \right)^{6} + \frac{1}{4} \right], & r \le 2^{1/6} d \\[1.5em] 0, & r > 2^{1/6} d \end{array} \right. ,
\end{align}
where $r$ is the minimal distance between a coupling point and the wall and $d$ is the `diameter' of the particle. Every coupling point interacts with the wall via the WCA potential, leading to an overall wall-swimmer interaction that models that of a hard rod or cylinder with a hard wall. We typically use $d = \sigma$.

\subsection{\label{sub:fig1}The Angular Evolution for Oscillating Swimmers}

\begin{figure}[]
  \centering
  \includegraphics[width=8.5 cm]{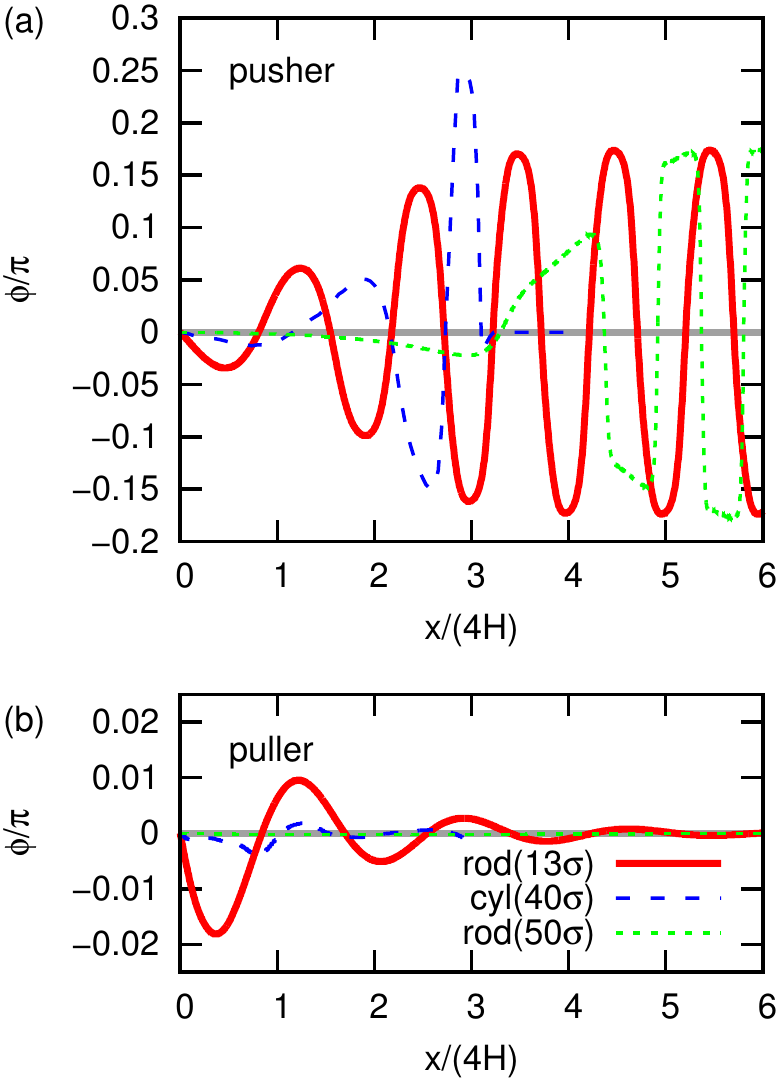}
  \caption{\label{fig:F1A}The angle $\phi$ as a function of horizontal displacement $x$ of swimmers between two parallel plates with separation $H$. Results are for swimmers that are initially oriented parallel to the walls and at $z=1\sigma$; using the exact same data sets as used in Fig.~1 of the main text. (a) The results for pushers: rod for $H=13\sigma$ (red, solid), cylinder for $H=40\sigma$ (blue, dashed), and rod for $H=50\sigma$ (green, dots). (b) The results for pullers, otherwise the systems are the same.}
\end{figure}

For completeness Fig.~\ref{fig:F1A} shows the way the angle $\phi$ evolves along the trajectory of the swimmers given in Fig.~1 of the main text. The orientation of the rod changes along the trajectory. When the particle moves between the two walls, it comes close to making a $45^{\circ}$ angle with respect to the horizontal. This is further visualized in the supplemental movies described in the next section.

\subsection{\label{sub:movies}Description of the Supplemental Movies}

To illustrate the movement of the swimmers, we have included two movies (not available in the arxiv version). These show the trajectory of a puller and pusher rod in a confining channel with height $H=13\sigma$ and lateral extent $L=70\sigma$. The initial position is $z=1\sigma$ and $\phi=0$ and we used a WCA parameter of $d = \sigma$. The labeling of the movies is as follows: the type of the particle is given, followed by a list of quantities and values, with each set separated by a double underscore. The notation of the quantities is the one used throughout and each quantity and value are separated by a single underscore. We chose a slightly smaller lateral extent of the channel than used to produce Fig.~1 of the main text. The reason is that for the typical channel sizes studied in our work, the motion of the swimmer would be difficult to observe. However, we have verified that the limited size of the channel does not strongly effect the trajectory.

\subsection{\label{sub:limit}LB Algorithm Limitations in the Near-wall Region}

We scrutinize the presence of the artificial limit cycle for our pusher-type rod through a series of computational examinations. By our examinations we reach the following conclusions. Since the LB algorithm does not explicitly account for near-wall lubrication corrections~\cite{fischer15}, it fails to be accurate in the near wall regime and we are therefore unable to comment on the nature of any potential limit cycle. Additionally, the counter-force point can artificially penetrate the wall at the point of closest approach. These points indicate that, although limit cycles may exist in certain physical swimmers, the simulated trajectories cannot offer physically relevant predictions. We explain the way we arrived at these conclusions in detail below.

\begin{figure}[]
  \centering
  \includegraphics[width=8.5 cm]{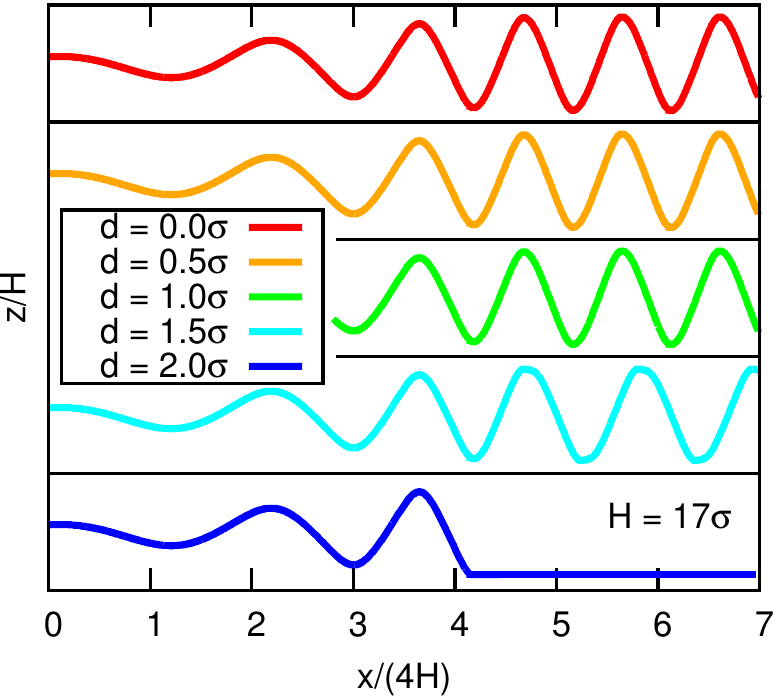}
  \caption{\label{fig:WCA}Trajectories of pusher-type rods between two parallel plates with separation $H = 17\sigma$. The horizontal displacement $x$ and vertical position $z$ are given for a swimmer that is initially oriented parallel to the walls $\phi = 0$ and at $z=1\sigma$, with $\sigma$ the MD unit of length and $z=0$ the center of the channel. For each curve, the top of the black frame enclosing the trajectory is at $z/H = 1/2$, while the bottom is at $z/H = -1/2$. From top to bottom, the range of the WCA interaction $d$ increases from 0 to $2\sigma$ in steps of $0.5\sigma$.}
\end{figure}

In our examination of the system, the lateral extent of the domain is varied between $L = 5H$ and $L = 35H$ to eliminate the effect of $xy$ periodicity on our results: there is no discernible impact of $L$ on the trajectories above $L = 10H$. We vary the viscosity and swimming speed to verify that retardation of the fluid momentum transport does not introduce these cycles; these changes only have a small effect. The value of the WCA interaction $d$ is varied, as shown explicitly in Fig.~\ref{fig:WCA}. We find that for $d > 1.5\sigma$ the limit cycle disappears and the rod's trajectory is reminiscent of the pusher cylinder's, see Fig. 1 in the main text. In both of these cases (inflated WCA rod and the unmodified cylinder) non-hydrodynamic contact with the WCA wall occurs and the self-propelled particles move along the plane of contact (sliding). Similar sliding dynamics have been observed in simulations that neglect HIs~\cite{Elgeti09}. 

The pusher rod performs its persistent oscillatory trajectory even in the absence of the WCA potential. Fortuitously, it does not penetrate the wall, although penetration can be achieved in this case by starting with values of $\phi$ that are greater than $\sim 25^{\circ}$ when $d = 0$. This may seem to indicate that the limit cycle is a physical effect. However, this is not the case, as the rod comes very close to the wall, where LB does not faithfully reproduce hydrodynamics~\cite{degraaf15b}.

\begin{figure}[]
  \centering
  \includegraphics[width=8.5 cm]{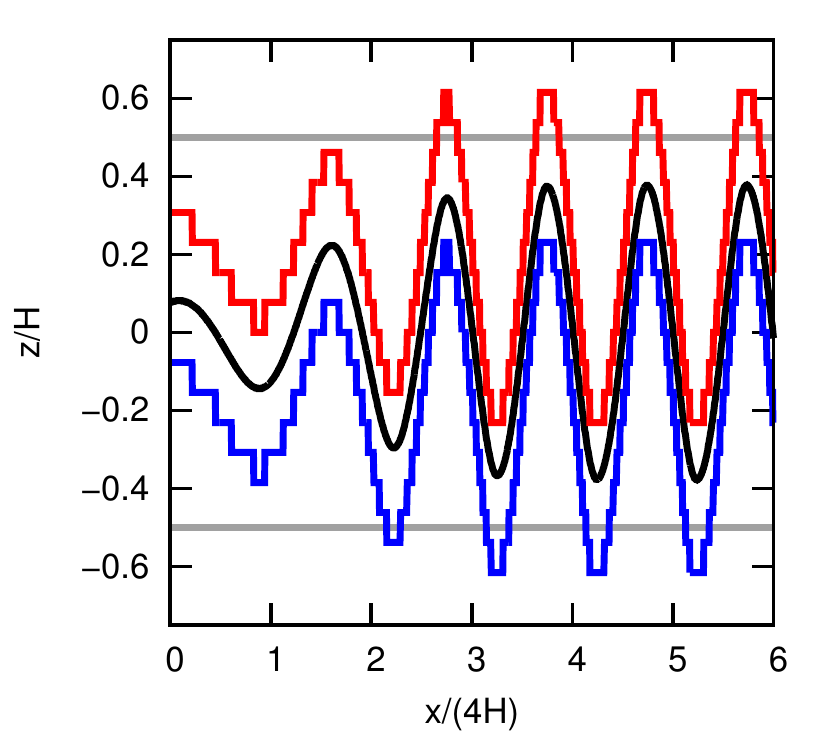}
  \caption{\label{fig:cfp}Trajectory (black curve) of a pusher-type rod between two parallel plates (gray lines) with separation $H = 13\sigma$. The horizontal displacement $x$ and vertical position $z$ are given for a swimmer that is initially oriented parallel to the walls $\phi = 0$ and at $z=1\sigma$, with $\sigma$ the MD unit of length and $z=0$ the center of the channel. The vertical extent of the lattice nodes involved in the 3-point interpolation of the counter-force are indicated using the red and blue lines.}
\end{figure}

We therefore also considered the interaction of the counter-force point with the wall, see Fig.~\ref{fig:cfp}. We find that the counter-force interpolation (which takes place over a region of $3\sigma$ in diameter due to the three-point coupling) is partially inside of the wall at the closest approach, which impacts the reorientation of the rod. To check the effect of this, we switched to a two-point interpolation stencil. The limit cycle persists, but here too the interpolation region overlaps with the wall nodes, even though the overlap is substantially reduced. When the value of $d$ increases beyond $d = 1.5\sigma$, the counter-force point is no longer interpolated inside the wall. Similarly, the cylinder's size prevents its counter-force point from being interpolated into the wall at closest approach. This indicates that the limit cycle observed for LB-raspberry swimmers is due to limitations in simulating the hydrodynamic interactions for close swimmer-wall separations, because of the spread-out counter-force scheme. 

While the near-wall hydrodynamics are not accurately captured by our algorithm, the far-field is. Therefore, in a system where there is a long-range (non-hydrodynamic) repulsion, our algorithm would produce the correct physics --- provided that the range of the repulsion is sufficient to keep the LB coupling points far enough away from the wall. In the main text, we chose the WCA repulsion in such a way that the size of the `hard core' matches the effective hydrodynamic size of the particle. Choosing the WCA range much larger, would remove this physical correspondence; therefore using an additional soft potential would be more appropriate to achieve wall repulsion. We are, however, unaware of any biological or artificial swimmers that are strongly repulsed from boundaries by long-ranged potentials and have therefore not considered this possibility further here.

In summary, the persistent oscillation (limit cycle) seen after long times for pusher-type swimmers must be attributed to a simulation artifact. Nevertheless, for the onset of the oscillation, which we are interested in the main text, there are no counter-force-overlap problems, as is clearly illustrated in Fig.~\ref{fig:cfp}.

\subsection{\label{sub:length}Rod Swimmer Length}

\begin{figure}[]
  \centering
  \includegraphics[width=8.5 cm]{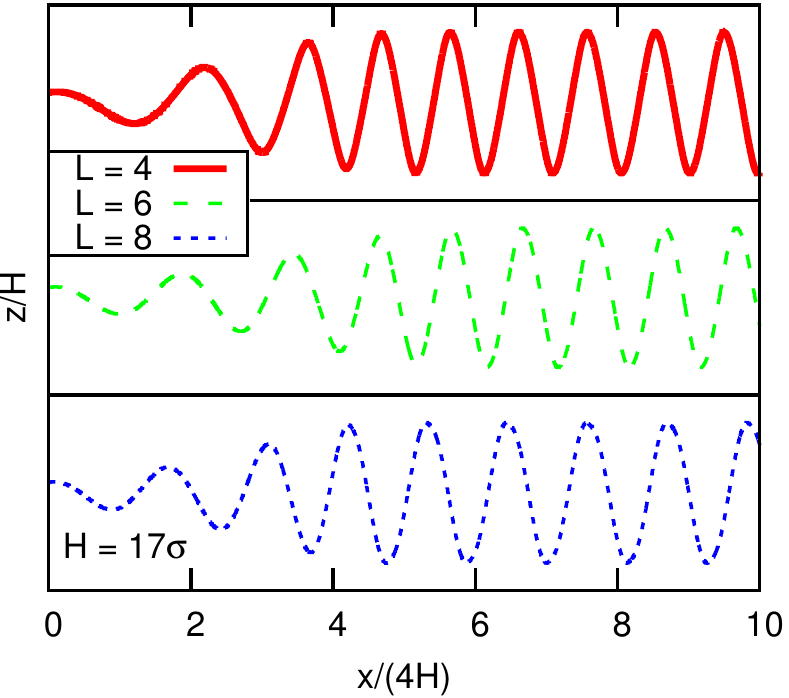}
  \caption{\label{fig:length}The ($x$,$z$) trajectories for pusher rods confined to a channel of height $H = 17\sigma$, which have initial position $z = 1\sigma$ and angle $\phi = 0$. We vary the length of the rod by adding coupling points. The rod used throughout has a bare length of $4\sigma$ (solid red curve) [and an effective hydrodynamic length of $5.5\sigma$]. Puller rods with bare length 6 (green dashed) and 8 (blue short dashes) are also shown. For each curve, the top of the black frame enclosing the trajectory is at $z/H = 1/2$, while the bottom is at $z/H = -1/2$.}
\end{figure}

The effect of rod length on the trajectories of pushers is seen in Fig.~\ref{fig:length}. Since the effective hydrodynamic diameter of the rods is governed primarily by the coupling parameters when using only a single row of coupling points~\cite{degraaf15a}, varying the length has the effect of varying the aspect ratio of rod-shaped particles. We found only minor modifications of the trajectories, reflecting the minor changes in the hydrodynamic multipole expansion due to the change in aspect ratio. That is, the presence of a hydrodynamic quadrupole is the dominant effect in the formation of oscillatory trajectories; the strength of the quadrupole moment is only weakly perturbed by the changes in the length that we considered.

\subsection{\label{sub:helical}Helical Trajectories}

\begin{figure}[]
  \centering
  \includegraphics[width=8.5 cm]{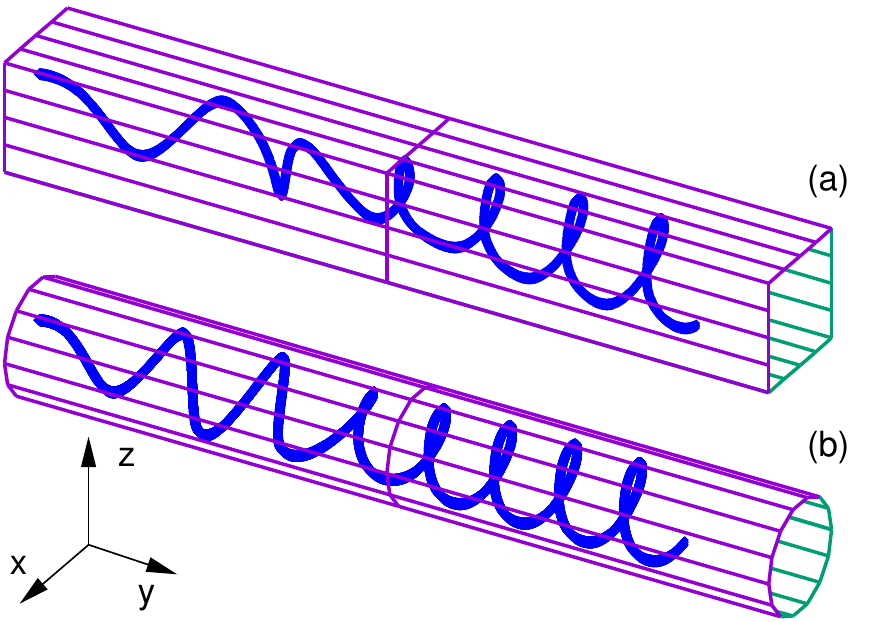}
  \caption{\label{fig:sqcyl}Trajectory (blue curve) of a pusher-type rod in a square tube (a) and a circular tube (b). The side length/diameter is $H = 17\sigma$. The swimmer is initially oriented parallel to the walls $\phi = 0$ and located at $x=2\sigma$ and $z=1\sigma$, with $\sigma$ the MD unit of length and $x=z=0$ the center of the tube. The trajectories shown here are $28H$ long in the $y$ direction.}
\end{figure}

The LB-raspberry swimmer model can be extended to simulate swimmers in other geometries. We find that our rods display helical trajectories, see Fig.~\ref{fig:sqcyl}. The helical trajectory observed for a tube with a square cross section is due to the swimmer starting off-center and away from one of the symmetry planes. Puller rods move consistently towards the center of the tube (not shown here) and also exhibit helical motion. The helical trajectory of the swimmer in the circular tube is due to a numerical artifact close to the boundary. The first part of the trajectory in the circular tube is purely oscillatory, as expected on the basis of symmetry and as we also observed for pullers. Only when an initial yaw angle is imposed does the rod perform a helical trajectory from the start of the simulation, similar to the observations of Ref.~\cite{zhu13}.

\section{\label{sec:flow}Swimmer-Generated Flow Fields}

As a swimmer at position $\vec{r}$ and orientation $\hat{p}$ moves, it disturbs the surrounding bulk fluid at position $\vec{x}$. This disturbance field can be written in terms of a multipole expansion
\begin{align}
  \label{eq:swimmerVelocityDefA}
  \vec{u}\left(\vec{x}, \vec{r}, \hat{p} \right) &= 
                        \kappa \vec{u}_\textmd{D} + 
                           \nu \vec{u}_\textmd{Q} + 
                          \mu \vec{u}_\textmd{SD} + 
                       o_1 \vec{u}_{\textmd{O}_1} + 
                       o_2 \vec{u}_{\textmd{O}_2} + 
                                             \ldots
\end{align}
where $\vec{u}_{D}$ is the Stokes dipole representing the opposing propulsion and drag forces, $\vec{u}_{Q}$ the quadrupole representing the fore-aft asymmetry of the swimmer, $\vec{u}_{SD}$ the source doublet representing the finite size of the swimmer, and the octupolar terms $\vec{u}_{\textmd{O}_1}$ and $\vec{u}_{\textmd{O}_2}$ describe the features of the flow in more detail \cite{mathijssen2015tracer}. Possible Stokeslet and rotlet (doublet) terms are omitted as the raspberry swimmers are force and torque free. 

These Newtonian flow fields can be written in terms of derivatives of the Oseen tensor
\begin{align}
  \label{eq:oseenTensor} \mathcal{J}_{ij}(\vec{r}) &= \frac{\delta_{ij}}{R} + \frac{R_i R_j}{R^3}, 
\end{align}
where $i,j \in \{x,y,z\}$ are indices. This produces a point-force Stokeslet $u_i^\textmd{S} = p_j \mathcal{J}_{ij}$ velocity field at position $\vec{x}$ due to a point force at position $\vec{r}$; where $\delta_{ij}$ is the Kronecker delta and $R = \left|\vec{x} - \vec{r}\right|$. From this \cite{mathijssen2015tracer} we obtain
\begin{align}
  \label{eq:flowStokesDipole}     \vec{u}_\textmd{D}(\vec{x}, \vec{r}, \hat{p})  &= + (\hat{p} \cdot \vec{\nabla}) \vec{u}^S, \\
  \label{eq:flowStokesQuadrupole} \vec{u}_\textmd{Q}(\vec{x}, \vec{r}, \hat{p})  &= - \tfrac{1}{2} (\hat{p} \cdot \vec{\nabla})^2 \vec{u}^S, \\
  \label{eq:flowSourceDoublet}    \vec{u}_\textmd{SD}(\vec{x}, \vec{r}, \hat{p}) &= - \tfrac{1}{2} \nabla^2 \vec{u}^S, \\
  \label{eq:flowStokesOctupole1}  \vec{u}_\textmd{O1}(\vec{x}, \vec{r}, \hat{p}) &= + \tfrac{1}{6} \nabla^2 (\hat{p} \cdot \vec{\nabla}) \vec{u}^S, \\
  \label{eq:flowStokesOctupole2}  \vec{u}_\textmd{O2}(\vec{x}, \vec{r}, \hat{p}) &= + \tfrac{1}{6} (\hat{p} \cdot \vec{\nabla})^3 \vec{u}^S,
\end{align}
where the derivatives act on the swimmer position $\vec{r}$. Hence, inserting equations (\ref{eq:flowStokesDipole}--\ref{eq:flowStokesOctupole2}) into \eqref{eq:swimmerVelocityDefA} gives the swimmer-generated flow field in the absence of walls.

\subsection{\label{sub:images}Wall-Induced Hydrodynamic Interactions}

These flow fields must be modified in the vicinity of boundaries. This can be done by the method of images so that the no-slip boundary conditions of both walls are satisfied. 
Consider the upper wall (denoted by superscript $+$): an additional velocity $\vec{u}^+$ must be added to the multipole expansion velocity $\vec{u}$ from (\ref{eq:swimmerVelocityDefA}) in order to satisfy the no-slip boundary condition at the wall $z=H/2$. We write this as $\left[ \vec{u} + \vec{u}^+ \right]_{z=H/2} = 0$. Likewise, the same is true at the bottom wall (denoted by superscript $-$) and we say $\left[\vec{u} + \vec{u}^- \right]_{z=-H/2} = 0$. The two velocity fields $\vec{u}^\pm$ represent an image system for the upper and lower wall, respectively. For two parallel walls the image system comprises an infinite series of images, but we consider only the first two images in this appendix. This derivation can be extended to $N$ images~\cite{mathijssen16}, and the results reported in the main text use eight images (four for each wall). 

These two images (above the upper wall [$^+$] and below the lower [$^-$]) are located at position 
\begin{align}
  \vec{r}^\pm &= \pm (H/2) \hat{e}_z + \underline{\underline{M}} (\mp (H/2) \hat{e}_z + \vec{r})
\end{align}
where $\underline{\underline{M}} = \mbox{diag}(1,1,-1)$. Hence, the relative distance between the images and a point in the fluid is $\vec{R}^\pm = \vec{x} - \vec{r}^\pm$. The velocity field of the image flow is then given by the Blake tensor in index notation
\begin{align}
  \label{eq:blakeTensor} 
  \mathcal{B}_{ij}^\pm &= \left(- \delta_{jk} + 2h^\pm \delta_{kz} \partial_j + (h^\pm)^2 M_{jk} \nabla^{2} \right) \mathcal{J}_{ik}, 
\end{align}
where $h^\pm = \tfrac{1}{2} (\vec{r}-\vec{r}^\pm)\cdot \hat{e}_z$, derivatives are taken with respect to swimmer position, and repeated indices are summed over. The two image systems due to point forces in the direction $\hat{p}$ are then $p_j \mathcal{B}_{ij}^\pm$. From this pair of Stokeslet images, the image systems of the Stokes dipole, quadrupole, \textit{etc.} can be constructed accordingly by taking successive derivatives as in equations (\ref{eq:flowStokesDipole}--\ref{eq:flowStokesOctupole2}) and the complete image system for the pair $\vec{u}^\pm$ is found. 

These image velocity fields interact hydrodynamically with the swimmer. The wall-induced translational and rotational velocities of the force-free and torque-free swimmer are found by rearranging the Fax{\'e}n relations evaluated at the swimmer position. Hence, we have
\begin{align}
  \label{eq:FaxenTranslationalVelocityFlow} \vec{v}_h &= \left[ \left(1+ \tfrac{1}{6}a^2  \nabla^2 \right) \vec{u}^\pm \right]_{\vec{x} = \vec{r}}, \\
  \label{eq:FaxenAngularVelocityFlow}  \vec{\Omega}_h &= \left[ 
  \tfrac{1}{2} \vec{\nabla} \times \vec{u}^\pm+ 
  G \hat{p} \times (\underline{\underline{E}}^\pm \cdot \hat{p}) \right]_{\vec{x} = \vec{r}},
\end{align}
where the derivatives act on the position $\vec{x}$, $\underline{\underline{E}}$ is the rate-of-strain tensor, $a$ is characteristic size of the swimmer, and $G = \tfrac{\gamma^2 - 1}{\gamma^2 + 1}$ is a function of the aspect ratio $\gamma$. Inserting the images of the swimmer-generated flow field \eqref{eq:swimmerVelocityDefA} into the Fax{\'e}n relations~(\ref{eq:FaxenTranslationalVelocityFlow}--\ref{eq:FaxenAngularVelocityFlow}) yields the wall-induced advection and rotation ($\vec{v}_\textmd{HI}$ and $\vec{\Omega}_\textmd{HI}$ in the equations of motion of the main text).

\section{\label{sub:dynamics}Swimmer dynamics model}

Using the translational invariance along the $x$ and $y$ directions, we write the swimmer's orientation as $\hat{p} = (\cos \phi, 0, \sin \phi)$ without loss of generality, where $\phi=0$ corresponds to swimming parallel to the walls. Hence, the swimmer's equations of motion simplify to the two coupled equations, $\dot{\phi} = \dot{\phi}(\phi, z)$ and $\dot{z} = \dot{z}(\phi,z)$. If we consider the simplified case of a point swimmer with aspect ratio $\gamma=1$, and only use one image system on each side of the channel, these equations are 
\begin{widetext}
\begin{small}
\begin{align}
  \label{eq:pdot}
  \dot{\phi} &=  
  \pm \frac{3 \kappa  \sin 2 \phi}{16 (z \mp \frac{H}{2})^3}
  \mp \frac{3 \nu (\cos \phi+3 \cos 3 \phi)}{64 (z \mp \frac{H}{2})^4}
  \mp \frac{3 \mu \cos \phi}{8 (z \mp \frac{H}{2})^4}
  \mp \frac{o_1 \sin 2 \phi}{4 (z \mp \frac{H}{2})^5}
  \mp \frac{3 o_2 (14 \sin 2 \phi+15 \sin 4 \phi)}{512 (z \mp \frac{H}{2})^5},\\
  \label{eq:zdot}
  \dot{z} &= 
  \pm \frac{3 \kappa  (3 \cos 2 \phi-1)}{16 (z \mp \frac{H}{2})^2}
  \pm \frac{\nu (\sin \phi+9 \sin 3 \phi)}{32 (z \mp \frac{H}{2})^3}
  \pm \frac{\mu \sin \phi}{(z \mp \frac{H}{2})^3}
  \mp \frac{5 o_1 (3 \cos 2 \phi-1)}{32 (z \mp \frac{H}{2})^4}
  \mp \frac{15 o_2 \cos ^2\phi (5 \cos 2 \phi-3)}{128 (z \mp \frac{H}{2})^4}
  + v_s \sin \phi .
\end{align}
\end{small}
\end{widetext}

\subsection{\label{sub:fitting}Fitting Hydrodynamic Moments}

An extension of these equations of motion (\ref{eq:pdot}--\ref{eq:zdot}), with $a \neq 0$ and $G \neq 0$, is used to match the dynamics of the model swimmers and LB swimmers (Table 1; main text). To achieve this, the time derivatives $\dot{\phi}$ and $\dot{z}$ are extracted from the LB trajectories for a number of randomly chosen $(\phi, z)$ coordinate points, $N=500$. Note that the first point in time is chosen to be after the first half oscillation such that the LB-raspberry swimmer has reached a constant swimming velocity and retardation effects are minimized. At each point, the LB values are compared to the values predicted by the model with a least squares method:
\begin{align}
  S = \sum_{i=1}^N \frac{(\dot{\phi}_\textmd{LB} - \dot{\phi}_\textmd{model})^2}{(2\pi)^2} + \frac{(\dot{z}_\textmd{LB} - \dot{z}_\textmd{model})^2}{H^2}.
\end{align}
Hence, the theory and LB simulations are matched by minimizing the $S$ function with respect to the far-field multipole expansion parameters. Here, the swimming speed $v_s$ is fixed at the actual values (Table 1; main text). Likewise, the particle radius $a$ is chosen to be fixed at the half-length of the LB rod or cylinder swimmer and the aspect ratio $\gamma$ is set to its geometric value. Similarly, the parameters $\mu, o_1$ are constrained to the LB-measured values, which is physically reasonable because these source doublets and quadrupolets are expected to be comparatively small, since our swimmers are constructed without fluid sources or sinks~\cite{degraaf16a}. Finally, the multipole moments $\kappa, \nu, o_2$ are allowed to vary, where a standard simulated annealing algorithm is used to find the least squares.

\subsection{\label{sub:oscillations}Analysis of swimmer oscillations}

In order to analyze the micro-swimmer dynamics, we linearize the equations of motion (\ref{eq:pdot}--\ref{eq:zdot}) about the centerline of the micro-channel ($z=0$), and about the orientation parallel to the walls ($\phi = 0$). For simplicity, we consider only the dipolar and quadrupolar contributions to the multipole expansion and set the octupolar and higher-order contributions to zero. The dynamics can then be captured by the matrix equation
\begin{align}
  \label{eq:matrixEOM}
  \left( \begin{array}{c} \dot{\phi}(t)  \\ \dot{z}(t)  \end{array} \right) &=
  \left(
  \begin{array}{cc}
  \displaystyle -\frac{6\kappa}{H^{3}} & 
  \displaystyle -\frac{48 (\nu + 2\mu)}{H^{5}}  \\[0.7em]
  \displaystyle v_s - \frac{2 (7 \nu+8\mu)}{H^{3}} &  
  \displaystyle \frac{12 \kappa}{H^{3}}  
  \end{array}
  \right) 
  \left( \begin{array}{c} \phi  \\ z  \end{array} \right).
\end{align}

First, we consider the motion in the absence of a dipole moment ($\kappa=0$), but with quadrupole moment $\nu$ and source doublet moment $\mu$. Then, the eigenvalues $\lambda_e$ of the matrix are
\begin{align}
  \lambda_e = \pm \frac{4 \sqrt{3(\nu +2 \mu ) \left(14 \nu +16 \mu - v_s H^3 \right)}}{H^4},
\end{align}
which corresponds to oscillatory motion ($\lambda_e$ is imaginary) if $H > \left( \left[14 \nu + 16 \mu\right] / v_s\right)^{1/3}$. That is to say that the channel must be wide enough with respect to $\nu$, $\mu$, and the swimming speed $v_s$ in order to observe oscillatory motion. For channels that are narrower than the critical height $H_c = \left( \left[14 \nu + 16 \mu\right] / v_s\right)^{1/3}$ the theory predicts that the swimmers do not oscillate. For our oscillating LB-raspberry swimmers this condition is met (Table 1; main text). Specifically, for the rod-type swimmer, we measured $\nu = 3.7~10^{-2} \sigma^{4}/ \tau$, $\mu = 0 \sigma^{4}/ \tau$ and $v_s = 2.5~10^{-3} \sigma/ \tau$. Therefore, condition for oscillatory motion is satisfied for channels heights $H_c \simeq 6$. Similarly for the cylinder-type LB swimmer, we find oscillatory motion requires $H_c \simeq 14$. In our simulations, we use channels heights that are larger than these critical values. 

Hence, oscillatory dynamics can be observed. With the initial conditions $z (0) = z_0$ and $\phi (0) = 0$, the swimmer's position in the channel is given by $z(t) \approx z_0 \cos(\omega t)$, where the oscillation frequency 
\begin{align}
  \omega &= i \lambda_e \approx 4 \sqrt{\frac{3 \nu v_s}{H^5}}
\end{align}
tends to zero as $\nu \to 0$ or $H \to \infty$, so that the oscillations gradually disappear in large channels.

With $\kappa$ included, the eigenvalues of equation \eqref{eq:matrixEOM} are
\begin{align}
  \lambda_e &= \frac{3 \kappa }{H^3} \pm i \omega \equiv \alpha \pm i \omega,\\
  \omega^2 &= \frac{48 v_s (\nu +2 \mu )}{H^5} - \frac{81 \kappa ^2}{H^6} - \frac{96 (\nu +2 \mu ) (7 \nu +8 \mu )}{H^8} ,
\end{align}
where we have introduced $\alpha = 3 \kappa/H^3$. Therefore, provided $\omega^2 >0$, the swimmer dynamics can be approximated by  
\begin{align}
  z(t) &\approx z_0 \cos(\omega t) \exp\left( \alpha t \right),
\end{align}
which describes oscillatory trajectories, growing in amplitude for pushers and decreasing for pullers.

\end{document}